\documentclass[pageno]{jpaper}

\usepackage[normalem]{ulem}
\usepackage{booktabs, tabularx}
\usepackage{enumitem}
\usepackage{amsfonts}
\usepackage[all]{nowidow}

\newcolumntype{R}{>{\raggedleft\arraybackslash}X}

\begin{document}

\title{Toward Practical Privacy-Preserving Convolutional Neural Networks\\ Exploiting Fully Homomorphic Encryption}

\author{
    Jaiyoung Park\textsuperscript{*}, Donghwan Kim\textsuperscript{*}, Jongmin Kim\textsuperscript{*}, Sangpyo Kim\textsuperscript{*}, Wonkyung Jung\textsuperscript{$\dagger$}, Jung Hee Cheon\textsuperscript{*}, and Jung Ho Ahn\textsuperscript{*}\\
    {\it\textsuperscript{*} Seoul National University, Seoul, South Korea}\\
    {\{{\it jeff1273, eastflame, jongmin.kim, vnb987, jhcheon, gajh}\}{\it @snu.ac.kr}}\\
    {\it \textsuperscript{$\dagger$} Samsung Electronics, Suwon, South Korea}\\
    {\it wk2.jung@samsung.com}
}
\date{}
\maketitle

\thispagestyle{empty}

\begin{abstract}
Incorporating fully homomorphic encryption (FHE) into the inference process of a convolutional neural network (CNN) draws enormous attention as a viable approach for achieving private inference (PI).
FHE allows delegating the entire computation process to the server while ensuring the confidentiality of sensitive client-side data.
However, practical FHE implementation of a CNN faces significant hurdles, primarily due to FHE's substantial computational and memory overhead.
To address these challenges, we propose a set of optimizations, which includes GPU/ASIC acceleration, an efficient activation function, and an optimized packing scheme.
We evaluate our method using the ResNet models on the CIFAR-10 and ImageNet datasets, achieving several orders of magnitude improvement compared to prior work and reducing the latency of the encrypted CNN inference to 1.4 seconds on an NVIDIA A100 GPU.
We also show that the latency drops to a mere 0.03 seconds with a custom hardware design.
\end{abstract}

\section{Introduction}
\label{sec:introduction}
Privacy regulations such as GDPR~\cite{law_2017_gdpr} have propelled data confidentiality to the forefront of concerns for cloud companies offering machine learning (ML) as a service.
Fully homomorphic encryption (FHE)~\cite{gentry_2009_fully}, which can evaluate arbitrary functions on encrypted data called ciphertext, has garnered attention as a promising solution with robust security assurance.
However, FHE-based computation exhibits distinct characteristics from their unencrypted counterpart, and na\"ively applying FHE to ML inference results in significant inefficiencies.
These disparities pose challenges in developing an end-to-end ML framework with FHE, which has been perceived to be costlier than alternative security solutions~\cite{huang_2022_cheetah,rathee_2020_cryptflow2}.

We unlock the potential of FHE-based solutions in the context of \emph{private inference} (PI).
Our focus lies on the framework for PI of convolutional neural networks (CNNs).
We employ the RNS-CKKS FHE scheme~\cite{cheon_18_rns}, which supports handling real and complex numbers.
The PI framework involves two parties: a client requesting CNN inference on encrypted data and a server conducting the evaluation using FHE operations.
The robust security offered by RNS-CKKS guarantees the non-disclosure of sensitive information, encompassing both the client's input data and the server's CNN model while securely delivering the inference results back to the client.

Our contributions span various layers of computer systems, and each contribution can be utilized separately to accelerate FHE-based CNN inference in other systems.
With all techniques combined, we achieve orders of magnitude higher performance for FHE-based CNN inference, reducing the encrypted inference latency of ResNet20 to a mere 1.4 seconds on a GPU and 0.03 seconds on a custom hardware design.

The following list summarizes the key contributions:

\begin{itemize} 
    \item We are unlocking new horizons in FHE-based PI by harnessing the latest GPU advancements and incorporating techniques derived from accelerator-based research, thereby introducing a new achievable milestone in this field.
    \item We combine AESPA~\cite{park_2022_aespa}, a novel activation function tailored to FHE-based PI, which converts ReLU into a simple quadratic function during inference through training-time specialization. 
    \item We leverage an efficient CNN library with novel packing methods named HyPHEN~\cite{kim_2023_hyphen}, which reduces the computation and memory requirements of FHE-based PI of CNNs.
\end{itemize} 

\section{A GPU Library Supporting RNS-CKKS}
\label{sec:gpu-lib}

Due to the high degree of parallelism in RNS-CKKS, GPUs have a great potential for performance enhancement by utilizing the massive number of computing resources in GPUs through parallelized computation.
In RNS-CKKS, the unit of computation is a polynomial in $\mathbb{Z}_Q[X]/(X^N + 1)$, which is an $L \times N$ integer matrix where $N$ is the degree of the polynomial and $L$ is the maximum multiplicative level of a ciphertext.
A polynomial is a huge matrix; typical values for $L$ and $N$ are respectively around $1$--$60$ and $2^{15}$--$2^{16}$.
By using GPUs, various computations on polynomials required for RNS-CKKS can mostly be performed in parallel.
For example, two polynomials can be added by $L\times N$ parallel element-wise additions.

However, the large size of the computational granularity also results in a memory bottleneck because faster memory units (e.g., L2 cache) in GPUs do not have enough capacity to accommodate it.
To overcome the bottleneck, we applied memory-centric optimizations for major computation kernels, such as NTT and base conversion, in prior work~\cite{jung_2021_100x, kim_2020_ntt}.
We have fused multiple GPU kernels to perform multiple jobs at once for each memory load and identified eligible RNS-CKKS parameter sets for GPUs (P2 and P3 in Table~\ref{tab:gpu-comparison} vs. P1).

Combining the latest GPU advancements and recent algorithmic optimizations, our GPU library outperforms state-of-the-art RNS-CKKS GPU studies, 100$\times$~\cite{jung_2021_100x} and TensorFHE~\cite{fan_2023_tensorfhe}, in major HE operations (see Table~\ref{tab:gpu-comparison}). Leveraging A100 leads to improved performance in homomorphic multiplication (HMult) compared to the previous use of V100 in 100$\times$. Algorithmic optimizations in bootstrapping (Boot), including techniques in \cite{bossuat_2021_efficient}, further widen the performance gap between our solution and the 100x bootstrapping.

\begin{table}[t]
\caption{Execution time (ms) comparison of our GPU library vs. state-of-the-art RNS-CKKS GPU acceleration studies, 100$\times$~\cite{jung_2021_100x} and TensorFHE~\cite{fan_2023_tensorfhe}. P1, P2, P3: fused, fused\textsubscript{L}, and fused\textsubscript{H} parameters in Table 2 of \cite{jung_2021_100x}. P4, P5: paramaters in Table 3 of \cite{jung_2021_100x}. }
\label{tab:gpu-comparison}
\begin{tabularx}{\columnwidth}{c|RRR}
\toprule
Impl. &\multicolumn{1}{c}{100$\times$} & \multicolumn{1}{c}{TensorFHE\textsuperscript{*}} & \multicolumn{1}{c}{Ours}  \\
\midrule
GPU &\multicolumn{1}{c}{V100} & \multicolumn{1}{c}{A100-SXM} & \multicolumn{1}{c}{A100-PCIe}\\
Word size &\multicolumn{1}{c}{64 bits} & \multicolumn{1}{c}{32 bits\textsuperscript{$\dagger$}} & \multicolumn{1}{c}{64 bits}\\
\midrule
HMult (P1) & 17.40 & 6.65 & 11.30\\
HMult (P2) & 2.96 & - & 2.59\\
HMult (P3) & 7.96 & - & 5.47\\
Boot (P4) & 328.25 & 250.45 & 171.27 \\
Boot (P5) & 526.96 & - & 355.84 \\
\bottomrule
\end{tabularx}
{
\footnotesize
\begin{itemize}[leftmargin=*, nolistsep, noitemsep]
\item[*] TensorFHE execution time is divided by 128 as it batches 128 operations.
\item[$\dagger$] We were unable to reproduce a working FHE implementation with the support for bootstrapping using the 32-bit word size due to the negative impact of small word sizes on the precision of RNS-CKKS. Recently proposed composite scaling~\cite{agrawal_2023_composite} can offset the precision loss, but it requires using 2$\times$ larger $L$ than that of 64-bit implementations for the same precision. Therefore, using the same parameter set greatly favors TensorFHE.
\end{itemize}
}
\vspace{0.05in}
\end{table}





\section{Replacing Activation Functions with AESPA}
FHE-based PI implementations are often bottlenecked by activation functions such as ReLU.
For example, \cite{lee_2022_low, lee_2021_precise} utilize a high-degree polynomial approximation of ReLU to implement RNS-CKKS CNN, but their implementation is severely bottlenecked by the high cost of evaluating a high-degree polynomial and frequent bootstrapping incurred by it; e.g., bootstrapping and ReLU account for 84\% of the total execution time in \cite{lee_2022_low}.

To remedy this problem, we exploit AESPA, a novel low-degree polynomial activation function that enables replacing ReLU with a quadratic function.
AESPA utilizes the orthogonal polynomial expansion of ReLU as an activation function and performs basis-wise normalization during training.
For brevity, we focus on the Hermite expansion of ReLU.
Let $\hat{f_i}$, $h_i$, and $d$ each denote the $i$-th coefficient for the Hermite expansion, the $i$-th Hermite polynomial, and the degree of Hermite expansion; our activation function is defined as follows:
\begin{equation}
    ReLU(x) = \gamma \sum_{i=0}^{d}\hat{f_i}\frac{h_i(x)-\mu}{\sqrt{\sigma^2+\epsilon}}+\beta = \gamma \sum_{i=0}^{d}\hat{f_i}\overline{h}_i(x)+\beta
\end{equation}

AESPA replaces both ReLU and batch normalization layers in a CNN.  
$\gamma$ and $\beta$ are learnable parameters and $\overline{h}_i(x)$ is a batch-normalized value of $h_i(x)$.
Please refer to \cite{park_2022_aespa} for a more detailed description of AESPA.

\section{HyPHEN: An Efficient Packing Method for FHE-based CNN}
ML frameworks such as PyTorch offer multiple memory formats dictating the data order of tensors (e.g., channel-last memory format).
Each format requires a distinct kernel to process the same operation.
This concept applies similarly in the context of FHE, where the memory format is analogous to \emph{packing} in FHE, although packing has more significant performance implications.
Numerous studies~\cite{aharoni_2023_helayers, LoLa, gilad_2016_cryptonets} have proposed different packing methods for FHE-based CNN, but they did not account for bootstrapping.
More recently, \cite{lee_2022_low} demonstrated an FHE-based CNN implementation with bootstrapping operations.
The packing method introduced in \cite{lee_2022_low} utilizes a single dense packing, which can effectively minimize the number of bootstrapping. 
However, their method suffers from the high cost of frequent homomorphic rotation operations, which are required to maintain the fixed packing method of \cite{lee_2022_low}, whereas input and output ciphertexts have different data orientations in FHE-based convolution layers; rotations account for more than 83\% of the total convolution time in \cite{lee_2022_low}.

\begin{figure}[t]
\centering
\includegraphics[width=0.9\columnwidth]{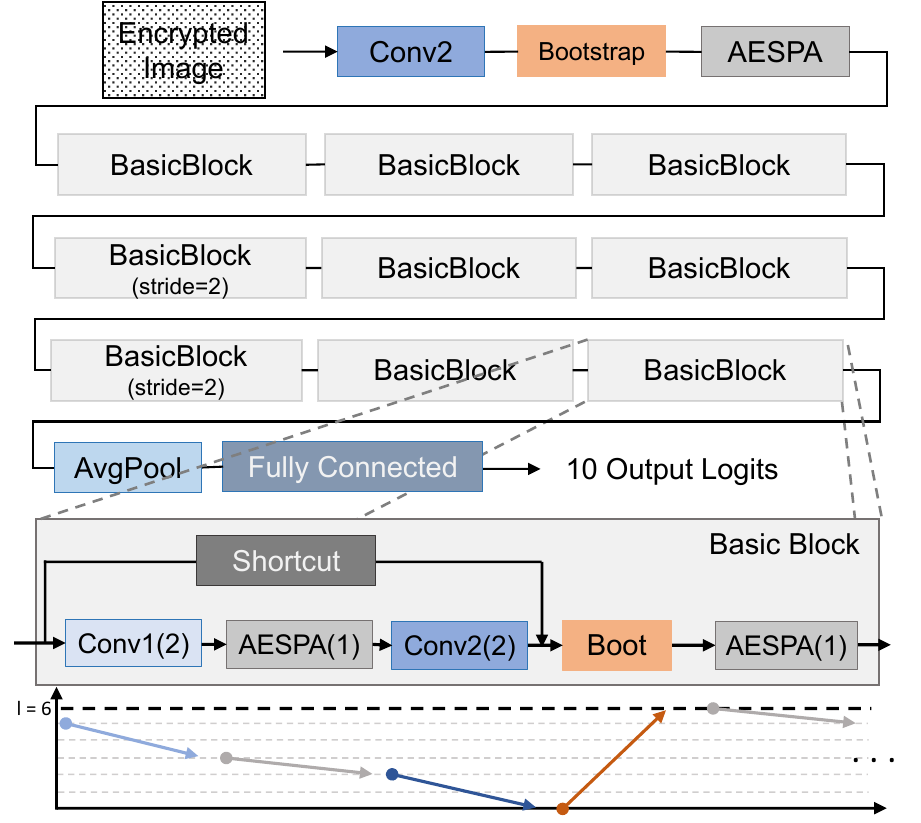} 
\caption{
ResNet20 model built with the HyPHEN basic blocks.
Each layer's multiplicative level consumption is also shown.
\label{fig:HyPHEN_architecture}}
\end{figure}

To mitigate this inefficiency, we designed HyPHEN, an FHE-based CNN library incorporating multiple packing methods.
By offering flexibility in the packing methods, we minimize the cost for rotations by choosing different packing methods before and after each convolution layer. 
Figure~\ref{fig:HyPHEN_architecture} depicts our construction of ResNet20.
Each basic block consists of two different convolution layers (each consuming two multiplicative levels) arranged in an interleaved manner with AESPA (each consuming one multiplicative level) activation functions.
We could minimize the cost of bootstrapping by placing the operation prior to the second activation function in a basic block, where the number of ciphertexts is small.
Please refer to \cite{kim_2023_hyphen} for a more detailed description of HyPHEN.

\section{Evaluation and Discussion}

We trained ResNet20 on the CIFAR10 dataset and ResNet18 on the ImageNet dataset using AESPA within PyTorch for both models.
CNN models trained with AESPA show comparable classification accuracies to the original networks with ReLU.
ResNet20 achieves an accuracy of 92.18\%, whereas the original accuracy of ResNet20 is 92.15\%.
Similarly, ResNet18 achieves an accuracy of 69.78\%, whereas the original model's accuracy is 69.75\%.

The latency of CNN inference for the two models is shown in Table~\ref{tab:latency_result}.
%
We reduce the inference latency of ResNet20 to mere 1.4 seconds, 1,622$\times$ faster than reported in \cite{lee_2022_low}.
Our GPU library reduces the execution time by 21.0$\times$ compared to that of our 64-thread CPU implementation.
Applying AESPA and HyPHEN results in an additional 6.12$\times$ speedup.
Meanwhile, our optimized ResNet18 inference takes 14.7 seconds.

\begin{table}[t]
\centering
\caption{Measured latency of FHE-based CNN inference on an NVIDIA A100 GPU and estimated latency on SHARP~\cite{kim_2023_sharp}, a state-of-the-art hardware accelerator proposal.}
\label{tab:latency_result}
\begin{tabularx}{\columnwidth}{l@{\hspace{0.05in}}RR@{\hspace{0.0in}}l@{\hspace{0.02in}}}
\toprule
& \multicolumn{2}{c}{\textbf{Latency (ms)}} \\
& \multicolumn{1}{c}{ResNet20} & \multicolumn{2}{c}{ResNet18}\\
\midrule
\cite{lee_2021_precise} (CPU 1-thread) & 10,602,000 & -\\
\cite{lee_2022_low} (CPU 1-thread) & 2,271,000 & -\\
\midrule
CPU 64-thread impl. of \cite{lee_2022_low} & 180,403 & 1,337,663 \\
GPU impl. of \cite{lee_2022_low} & 8,575 & 76,503 & \textsuperscript{*}\\      
AESPA + HyPHEN (GPU) & 1,402 & 14,690 \\
\midrule
\cite{lee_2022_low} (SHARP) & 99 & - \\
AESPA + HyPHEN (SHARP) & 30 & - \\
\bottomrule
\end{tabularx}
{
\footnotesize
\begin{itemize}[leftmargin=*, nolistsep, noitemsep]
\item[*] Due to the limited GPU memory capacity, we could not run ResNet18 directly on the GPU with \cite{lee_2022_low}.
To estimate its performance, we assumed that the GPU can contain the entire working set, ignoring the host-to-GPU data movement cost.
Meanwhile, AESPA + HyPHEN is free from this problem due to optimizations that effectively reduce the working set size.
\end{itemize}
}
\end{table}

There still exists a vast room for further improvement in FHE-based CNN performance.
In particular, numerous specialized hardware accelerator designs~\cite{kim_2023_sharp, kim_2022_ark,kim_2022_bts,samardzic_2021_f1,samardzic_2022_craterlake}
have been proposed recently.
We estimated the performance of our CNN implementation on a state-of-the-art hardware accelerator proposal, SHARP~\cite{kim_2023_sharp}, through simulation (see Table~\ref{tab:latency_result}).
Simulation results show that specialized hardware enables real-time FHE-based CNN inference by reducing the inference time to as low as 30 milliseconds, 75,700$\times$ faster than the original single-threaded CPU implementation.


\bibliographystyle{IEEEtranS}
\bibliography{reference}

\begin{thebibliography}{10}
\providecommand{\url}[1]{#1}
\csname url@samestyle\endcsname
\providecommand{\newblock}{\relax}
\providecommand{\bibinfo}[2]{#2}
\providecommand{\BIBentrySTDinterwordspacing}{\spaceskip=0pt\relax}
\providecommand{\BIBentryALTinterwordstretchfactor}{4}
\providecommand{\BIBentryALTinterwordspacing}{\spaceskip=\fontdimen2\font plus
\BIBentryALTinterwordstretchfactor\fontdimen3\font minus \fontdimen4\font\relax}
\providecommand{\BIBforeignlanguage}[2]{{%
\expandafter\ifx\csname l@#1\endcsname\relax
\typeout{** WARNING: IEEEtranS.bst: No hyphenation pattern has been}%
\typeout{** loaded for the language `#1'. Using the pattern for}%
\typeout{** the default language instead.}%
\else
\language=\csname l@#1\endcsname
\fi
#2}}
\providecommand{\BIBdecl}{\relax}
\BIBdecl

\bibitem{agrawal_2023_composite}
R.~Agrawal, J.~Ahn, F.~Bergamaschi, R.~Cammarota, J.~H. Cheon, F.~D.~M. de~Souza, H.~Gong, M.~Kang, D.~Kim, J.~Kim, H.~de~Lassus, J.~H. Park, M.~Steiner, and W.~Wang, ``{High-precision RNS-CKKS on Fixed but Smaller Word-size Architectures: Theory and Application},'' Cryptology ePrint Archive, Paper 2023/1462, 2023.

\bibitem{aharoni_2023_helayers}
E.~Aharoni, A.~Adir, M.~Baruch, N.~Drucker, G.~Ezov, A.~Farkash, L.~Greenberg, R.~Masalha, G.~Moshkowich, D.~Murik, H.~Shaul, and O.~Soceanu, ``{HeLayers: A Tile Tensors Framework for Large Neural Networks on Encrypted Data},'' in \emph{Privacy Enhancing Technologies Symposium (PETS)}, 2023.

\bibitem{bossuat_2021_efficient}
J.~Bossuat, C.~Mouchet, J.~R. Troncoso{-}Pastoriza, and J.~Hubaux, ``{Efficient Bootstrapping for Approximate Homomorphic Encryption with Non-sparse Keys},'' in \emph{Annual International Conference on the Theory and Applications of Cryptographic Techniques (Eurocrypt)}, 2021.

\bibitem{LoLa}
A.~Brutzkus, R.~Gilad{-}Bachrach, and O.~Elisha, ``{Low Latency Privacy Preserving Inference},'' in \emph{International Conference on Machine Learning (ICML)}, 2019.

\bibitem{cheon_18_rns}
J.~H. Cheon, K.~Han, A.~Kim, M.~Kim, and Y.~Song, ``{A Full {RNS} Variant of Approximate Homomorphic Encryption},'' in \emph{Selected Areas in Cryptography (SAC)}, 2018.

\bibitem{fan_2023_tensorfhe}
S.~Fan, Z.~Wang, W.~Xu, R.~Hou, D.~Meng, and M.~Zhang, ``{TensorFHE: Achieving Practical Computation on Encrypted Data Using GPGPU},'' in \emph{IEEE International Symposium on High-Performance Computer Architecture (HPCA)}, 2023.

\bibitem{gentry_2009_fully}
C.~Gentry, ``{Fully Homomorphic Encryption Using Ideal Lattices},'' in \emph{Annual ACM Symposium on Theory of Computing (STOC)}, 2009.

\bibitem{gilad_2016_cryptonets}
R.~Gilad{-}Bachrach, N.~Dowlin, K.~Laine, K.~E. Lauter, M.~Naehrig, and J.~Wernsing, ``{CryptoNets: Applying Neural Networks to Encrypted Data with High Throughput and Accuracy},'' in \emph{International Conference on Machine Learning (ICML)}, 2016.

\bibitem{huang_2022_cheetah}
Z.~Huang, W.~jie Lu, C.~Hong, and J.~Ding, ``{Cheetah: Lean and Fast Secure {Two-Party} Deep Neural Network Inference},'' in \emph{USENIX Security Symposium}, 2022.

\bibitem{jung_2021_100x}
W.~Jung, S.~Kim, J.~Ahn, J.~H. Cheon, and Y.~Lee, ``{Over 100x Faster Bootstrapping in Fully Homomorphic Encryption through Memory-centric Optimization with GPUs},'' \emph{IACR Transactions on Cryptographic Hardware and Embedded Systems (TCHES)}, vol. 2021, no.~4, 2021.

\bibitem{kim_2023_hyphen}
D.~Kim, J.~Park, J.~Kim, S.~Kim, and J.~Ahn, ``{HyPHEN: A Hybrid Packing Method and Optimizations for Homomorphic Encryption-Based Neural Networks},'' arXiv preprint arXiv:2302.02407, 2023.

\bibitem{kim_2023_sharp}
J.~Kim, S.~Kim, J.~Choi, J.~Park, D.~Kim, and J.~Ahn, ``{SHARP: A Short-Word Hierarchical Accelerator for Robust and Practical Fully Homomorphic Encryption},'' in \emph{Annual International Symposium on Computer Architecture ({ISCA})}, 2023.

\bibitem{kim_2022_ark}
J.~Kim, G.~Lee, S.~Kim, G.~Sohn, M.~Rhu, J.~Kim, and J.~Ahn, ``{ARK: Fully Homomorphic Encryption Accelerator with Runtime Data Generation and Inter-Operation Key Reuse},'' in \emph{{IEEE/ACM International Symposium on Microarchitecture (MICRO)}}, 2022.

\bibitem{kim_2020_ntt}
S.~Kim, W.~Jung, J.~Park, and J.~Ahn, ``{Accelerating Number Theoretic Transformations for Bootstrappable Homomorphic Encryption on GPUs},'' in \emph{IEEE International Symposium on Workload Characterization (IISWC)}, 2020.

\bibitem{kim_2022_bts}
S.~Kim, J.~Kim, M.~J. Kim, W.~Jung, J.~Kim, M.~Rhu, and J.~Ahn, ``{BTS: An Accelerator for Bootstrappable Fully Homomorphic Encryption},'' in \emph{Annual International Symposium on Computer Architecture (ISCA)}, 2022.

\bibitem{lee_2022_low}
E.~Lee, J.~Lee, J.~Lee, Y.~Kim, Y.~Kim, J.~No, and W.~Choi, ``{Low-Complexity Deep Convolutional Neural Networks on Fully Homomorphic Encryption Using Multiplexed Parallel Convolutions},'' in \emph{International Conference on Machine Learning (ICML)}, 2022.

\bibitem{lee_2021_precise}
J.~Lee, E.~Lee, J.~Lee, Y.~Kim, Y.~Kim, and J.~No, ``{Precise Approximation of Convolutional Neural Networks for Homomorphically Encrypted Data},'' \emph{IEEE Access}, vol.~11, 2023.

\bibitem{park_2022_aespa}
J.~Park, M.~J. Kim, W.~Jung, and J.~Ahn, ``{{AESPA:} Accuracy Preserving Low-degree Polynomial Activation for Fast Private Inference},'' arXiv preprint arXiv:2201.06699, 2022.

\bibitem{rathee_2020_cryptflow2}
D.~Rathee, M.~Rathee, N.~Kumar, N.~Chandran, D.~Gupta, A.~Rastogi, and R.~Sharma, ``{CrypTFlow2: Practical 2-Party Secure Inference},'' in \emph{{ACM} {SIGSAC} Conference on Computer and Communications Security (CCS)}, 2020.

\bibitem{samardzic_2021_f1}
N.~Samardzic, A.~Feldmann, A.~Krastev, S.~Devadas, R.~Dreslinski, C.~Peikert, and D.~Sanchez, ``{F1: A Fast and Programmable Accelerator for Fully Homomorphic Encryption},'' in \emph{IEEE/ACM International Symposium on Microarchitecture (MICRO)}, 2021.

\bibitem{samardzic_2022_craterlake}
N.~Samardzic, A.~Feldmann, A.~Krastev, N.~Manohar, N.~Genise, S.~Devadas, K.~Eldefrawy, C.~Peikert, and D.~Sanchez, ``{CraterLake: A Hardware Accelerator for Efficient Unbounded Computation on Encrypted Data},'' in \emph{Annual International Symposium on Computer Architecture (ISCA)}, 2022.

\bibitem{law_2017_gdpr}
P.~Voigt and A.~Von~dem Bussche, \emph{{The EU General Data Protection Regulation (GDPR)}}.\hskip 1em plus 0.5em minus 0.4em\relax Springer, 2017.

\end{thebibliography}

\end{document}